\documentclass[aps,prl,twocolumn,superscriptaddress,preprintnumbers,longbibliography]{revtex4-2}

\usepackage{amsmath}
\usepackage{graphicx}
\usepackage{epstopdf}
\usepackage{xcolor}

\newcommand{\be}{\begin{equation}}
\newcommand{\ee}{\end{equation}}
\newcommand{\bea}{\begin{eqnarray}}
\newcommand{\eea}{\end{eqnarray}}
\newcommand{\nn}{\nonumber}

\begin{document}

\title{Quadratic-in-spin interactions at fifth post-Newtonian order probe new physics}

\author{Jung-Wook Kim}
\affiliation{Queen Mary University of London, London E1 4NS, United Kingdom}

\author{Mich\`ele Levi}
\email{levi@maths.ox.ac.uk}
\affiliation{University of Oxford, Oxford OX2 6GG, United Kingdom}
\affiliation{Queen Mary University of London, London E1 4NS, United Kingdom}
\affiliation{Niels Bohr Institute, University of Copenhagen, 2100 Copenhagen, Denmark}

\author{Zhewei Yin}
\affiliation{Uppsala University, 75108 Uppsala, Sweden}

\date{\today}

\begin{abstract}
We obtain for the first time all quadratic-in-spin interactions in spinning binaries at the third 
subleading order in post-Newtonian (PN) gravity, and provide their observable binding energies 
and their gauge-invariant relations to the angular momentum.
Our results are valid for generic compact objects, orbits, and spin orientations, and enter at the 
fifth PN order for maximally-rotating objects, thus pushing the state of the art. 
This is accomplished through an extension of the effective field theory of spinning gravitating 
objects, and of its computational application.
We also discover a new finite-size effect 
which is unique to spinning objects, with a new ``Spin Love number'' as its characteristic coefficient, 
that is a new probe for gravity and QCD. 
\end{abstract}

\preprint{QMUL-PH-21-52, UUITP-61/21}

\maketitle

The success of ground-based experiments in measuring gravitational waves (GWs) since the first 
detection from a black-hole (BH) binary merger \cite{Abbott:2016blz} by the Advanced LIGO 
\cite{LIGOScientific:2014pky} and Advanced VIRGO \cite{VIRGO:2014yos} collaboration has 
exceeded most expectations. 
By now there is a worldwide network of GW detectors of second-generation technology, as the 
twin Advanced LIGO detectors in the US have been joined by the Advanced Virgo detector in 
Europe \cite{VIRGO:2014yos}, and then by the KAGRA detector in Japan \cite{KAGRA:2020tym}. 
These experiments have been continually reaching higher sensitivities, which yield more frequent 
detections, and the influx of data has been steeply growing 
\cite{LIGOScientific:2018mvr,LIGOScientific:2020ibl,LIGOScientific:2021djp}. 

Since 2017 these detections also include neutron stars (NSs) as individual components of the binaries, in 
NS binaries \cite{LIGOScientific:2017vwq} or mixed NS-BH binaries \cite{LIGOScientific:2021qlt}. The study 
of these various GW sources in the inspiral phase, when the components of the binary are still orbiting in 
non-relativistic velocities, is carried out analytically via the post-Newtonian (PN) approximation of 
General Relativity (GR) \cite{Blanchet:2013haa}. This enables to model theoretical gravitational waveforms, 
which inform us on a wealth of astrophysical and cosmological scenarios that were previously unthought-of, 
and uniquely also on gravity in the strong-field regime, and QCD in extreme conditions for NSs 
\cite{TheLIGOScientific:2016src, TheLIGOScientific:2016wfe}. To this end, it is crucial to study spin and 
finite-size effects, as all components of the binaries are in fact spinning gravitating objects 
\cite{Abbott:2016izl}.

In both traditional GR and modern HEP approaches to study these GW sources, the compact objects that 
make up the binaries are essentially captured by an effective description of a point particle that is 
endowed with characteristic coefficients which encapsulate the physics at the small scales of its internal 
structure \cite{Goldberger:2004jt,Goldberger:2007hy,Levi:2018nxp}. These effective characteristic 
coefficients are generally referred to as ``Wilson coefficients'' in the language of effective field theory 
(EFT) from QFT. Determining the numerical values of these coefficients in the low-energy or large-scale 
approximation constitutes the final and often most challenging piece of fixing the effective theory. This 
task, commonly referred to  as ``matching'' in EFT parlance, is tackled in an indefinite variety of ways, 
e.g.~via analytical studies of specific observables in the full theory, numerical simulations of the full 
theory, or by simply matching the unknown coefficients to experimental data.

In the non-spinning case the effective description of a point particle remains trivial till high PN orders, 
where the simple point-mass alone is sufficient up to the fifth PN (5PN) order, that has been approached 
only recently after decades of studies in PN theory. Finite-size effects thus enter only at the 5PN 
order in this simplified case, preceded by effective coefficients that correspond to the so-called ``Love 
numbers''. These have been introduced more than a century ago in Newtonian theory for planetary bodies as 
the parameters that measure their rigidity, and thus their response to tidal forces. In the context of 
BH physics the related numbers have been studied for almost 4 decades already, see 
e.g.~\cite{Binnington:2009bb} for reference, where general studies in GR have been carried out mainly in 
the last 15 years, pioneered by e.g.~\cite{Hinderer:2007mb,Damour:2009vw,Binnington:2009bb}. 

For the real spinning case the physics gets dramatically more complicated. To begin with, the spin induces 
higher multipoles to all orders, and the associated finite-size effects enter already as of the 2PN order 
with the spin-induced quadrupole \cite{Barker:1975ae}. These finite-size effects are characterized by 
coefficients commonly referred to in GR as ``multipole deformation parameters'', see 
e.g.~\cite{Poisson:1997ha}, corresponding to analogous Wilson coefficients in the EFT description 
\cite{Levi:2015msa}, see \eqref{spinleadingnmc} below. The ``multipole deformation parameters'' are not to 
be confused with the aforementioned ``Love numbers'', see \eqref{ordinaryLove} below. For example, whereas 
``Love numbers'', which can also be studied for spinning objects, see e.g.~a recent surge of studies 
\cite{LeTiec:2020spy,LeTiec:2020bos,Chia:2020yla,Charalambous:2021mea,Charalambous:2021kcz,Castro:2021wyc}, 
have been shown in virtually all studies to date, to vanish for BHs in GR (in 4 dimensions), the 
spin-induced ``multipole deformation parameters'' in contrast equal 1 for BHs 
\cite{Levi:2015msa,Levi:2018nxp}. 

In recent years impressive progress has been made in the state of the art of PN theory for the conservative 
dynamics of an inspiraling binary. In particular, the point-mass interaction at the 5PN order has been 
recently accomplished via a combined exploitation of traditional GR methods 
\cite{Bini:2019nra,Bini:2020wpo,Bini:2020uiq}, with crucial ingredients taken from self-force theory, and 
the effective-one-body approach \cite{Buonanno:1998gg}. Shortly after, this sector was also confirmed via 
an EFT computation \cite{Blumlein:2020pyo}. However, in order to attain any PN accuracy (beyond 1PN), the 
spinning case must be tackled. \cite{Antonelli:2020aeb,Antonelli:2020ybz} then followed the footsteps of 
\cite{Bini:2019nra,Bini:2020wpo} in implementing a similar approach to the sector that is linear in 
the spins at 4.5PN order, and for a limited simplified configuration of circular orbits with aligned spins 
-- to the piece that is linear in the spins at 5PN order \cite{Antonelli:2020ybz}.

It is critical to note however that while it is important and illustrative to target specific new PN 
sectors via the capitalization on available results from existing elementary methods, such an ad-hoc 
approach is essentially limited. It does not provide a conceptual framework to generally tackle the various 
sectors required to a certain accuracy, nor does it provide an independent framework to study PN theory, 
and thus it is also prone to the propagation of errors from the combined inputs of the various ingredient 
methods.  

In this letter we tackle the 5PN order with spins in the most generic settings, obtaining for the first 
time all the interactions that are quadratic in the spins. Our derivation builds on the EFT of spinning 
gravitating objects introduced in \cite{Levi:2015msa}, see also \cite{Levi:2018nxp,Levi:2017kzq}, and its 
extensions \cite{Levi:2011eq, Levi:2014sba, Levi:2014gsa, Levi:2015uxa, Levi:2015ixa, Levi:2016ofk, 
Levi:2019kgk,Levi:2020kvb,Levi:2020uwu,Levi:2020lfn}. It is the most formidable undertaking in PN 
theory with spins as yet. The cutting-edge calculation of the present sectors outlined here also 
serves as a unique computational experiment, to eventually discover a new feature in the theory of a 
spinning particle: a new type of finite-size effect, which is \textit{unique to spinning objects}, with  
new ``Spin Love numbers''. These do not exist in the non-spinning case, and thus they provide a new 
unique probe for gravity and QCD.

\paragraph{EFT of spinning gravitating objects.} 

We build on the EFT of spinning gravitating objects introduced in \cite{Levi:2015msa}. To obtain all the 
quadratic-in-spin interactions, we need to start from the two-particle effective action for the compact 
binary \cite{Goldberger:2004jt,Levi:2018nxp}:
\be \label{2ptact}
S_{\text{eff}}=S_{\text{g}}[g_{\mu\nu}(x)]+\sum_{a=1}^{2}S_{\text{pp}}[(\lambda_a)],
\ee
and then carefully consider the effective action of the spinning particle, $S_{\text{pp}}$, localized 
on the worldline parametrized by $\lambda_a$ for each of the two components of the binary. 

First, in the non-relativistic approximation it is useful to employ a Kaluza-Klein time+space decomposition 
of the field \cite{Kol:2007bc,Kol:2010ze}, which was first tested in sectors with spins in 
\cite{Levi:2008nh,Levi:2010zu}.
The spatial dimension, $d$, must be kept generic throughout, as dimensional regularization will be used to 
evaluate the Feynman integrals, with the modified minimal subtraction ($\overline{\text{MS}}$) prescription 
applied through the $d$-dimensional gravitational constant \cite{Levi:2020kvb}:
\be 
\label{msbar}
G_d\equiv G_N \left(\sqrt{4\pi e^{\gamma_E}} \,R_0 \right)^{d-3},
\ee
in which $G_N\equiv G$ is Newton's gravitational constant in three-dimensional space, 
$\gamma_E$ 
is Euler's constant, and $R_0$ is some fixed renormalization scale. 

The quadratic-in-spin sectors include finite-size effects in addition to the minimal coupling of spinning 
objects to gravity, so we need to consider the following extended effective action for each of the two 
spinning particles \cite{Levi:2015msa,Levi:2018nxp}:
\begin{align} 
\label{spinptact}
S_{\text{pp}}[(\lambda)] =\int 
d\lambda \bigg[-m \sqrt{u^2}- &\frac{1}{2} \hat{S}_{\mu\nu} \hat{\Omega}^{\mu\nu}
-\frac{\hat{S}^{\mu\nu} p_{\nu}}{p^2} \frac{D p_{\mu}}{D \lambda} \nn\\
 & + L_{\text{NMC}}
\Big[g_{\mu\nu},u^{\mu},S^{\mu}\Big]\bigg],
\end{align}
where the non-minimal coupling of gravity to spin, $L_{\text{NMC}}$, is formulated in terms of the 
definite-parity classical analogue of the Pauli-Lubanski pseudovector, $S_{\mu}$, as defined in 
\cite{Levi:2014gsa,Levi:2015msa,Levi:2019kgk}. We note that another treatment of spin utilizing EFT 
techniques was approached in \cite{Porto:2005ac}, where it was applied to low PN orders.

The non-minimal coupling of gravity to all orders in spin, that is linear in the curvature, is given in the 
following compact form \cite{Levi:2015msa}:
\begin{align}
\label{spinleadingnmc}
L_{\text{NMC(R)}}&
=\sum_{n=1}^\infty \frac{(-1)^n}{(2n)!}\frac{C_{ES^{2n}}}{m^{2n-1}}
D_{\mu_{2n}}\cdots D_{\mu_{3}}\frac{E_{\mu_{1}\mu_{2}}}{\sqrt{u^2}}\nn\\
& \qquad \qquad \qquad \quad \bullet S^{\mu_1}S^{\mu_2}\cdots S^{\mu_{2n-1}}S^{\mu_{2n}}\nn\\
&+\sum_{n=1}^\infty \frac{(-1)^n}{(2n+1)!}\frac{C_{BS^{2n+1}}}{m^{2n}}
D_{\mu_{2n+1}}\cdots D_{\mu_{3}}\frac{B_{\mu_{1}\mu_{2}}}{\sqrt{u^2}}\nn\\
& \qquad \qquad \qquad \quad \bullet S^{\mu_1}S^{\mu_2}\cdots S^{\mu_{2n}}S^{\mu_{2n+1}},
\end{align}
with the definite-parity electric and magnetic components of the curvature:
\bea
E_{\mu\nu}&\equiv& R_{\mu\alpha\nu\beta}u^{\alpha}u^{\beta}, \label{elec}\\
B_{\mu\nu}&\equiv& \frac{1}{2} \epsilon_{\alpha\beta\gamma\mu} 
R^{\alpha\beta}_{\,\,\,\,\,\,\,\delta\nu}u^{\gamma}u^{\delta}\label{mag},
\eea
and their covariant derivatives, $D_{\mu}$. In this infinite series we introduced an infinite set of Wilson 
coefficients, which correspond to the aforementioned ``multipole deformation parameters''. The only term 
from this series that contributes to our present sectors is the first electric term preceded by a Wilson 
coefficient, which corresponds to the quadrupolar deformation constant, similar to \cite{Barker:1975ae}.

Yet, at this high PN order the effective action of a spinning particle needs to be extended beyond linear 
order in the curvature, namely beyond \eqref{spinleadingnmc}. Such extension was briefly suggested very 
recently in \cite{Levi:2020uwu} from basic symmetry considerations and reasoning 
\cite{Levi:2015msa,Levi:2018nxp}. From dimensional analysis and power counting only the following 
terms may enter at the 5PN order to quadratic order in the spins:
\begin{subequations}
\label{nmcspinRR}
\begin{align}
L_{\text{NMC(R$^2$)}}&
= C_{E^2} \frac{E_{\alpha\beta}E^{\alpha\beta}}{\sqrt{u^2}^{\,3}}
+ C_{B^2} \frac{B_{\alpha\beta}B^{\alpha\beta}}{\sqrt{u^2}^{\,3}}
\label{ordinaryLove}\\
+ \,C_{E^2S^2} & S^{\mu} S^{\nu} \frac{E_{\mu\alpha}E_{\nu}^{\,\alpha}}{\sqrt{u^2}^{\,3}}
+ C_{B^2S^2} S^{\mu} S^{\nu} \frac{B_{\mu\alpha}B_{\nu}^{\alpha}}{\sqrt{u^2}^{\,3}}.
\label{SpinLove}
\end{align}
\end{subequations}
Interestingly, non-minimal couplings that are only linear in the spins, enter only at higher PN orders. 
The terms in \eqref{nmcspinRR} involve coefficients that at this point absorb all numerical and mass
factors (unlike those in \eqref{spinleadingnmc}). 
Yet further scrutiny of \eqref{ordinaryLove}, \eqref{SpinLove} reveals that their dimensionless Wilson 
coefficients should be defined as 
\begin{align}
\label{scalingWilson} 
C_{[E/B]^2} &\to + \frac{1}{2} G^4 m^5 \,C_{[E/B]^2},\\
C_{[E/B]^2S^2} & \to + \frac{1}{2} G^2 m \,\, C_{[E/B]^2S^2}.
\label{scalingWilsonS} 
\end{align}
Notably these are the first terms that exhibit an additional scaling in $G$ in their coefficients. This is 
unlike any PN contributions previously encountered,  
where the order in $G$ of the leading field couplings has always been identical to the order of their 
overall contributions. 

The coefficients in \eqref{ordinaryLove}, \eqref{scalingWilson} correspond to the aforementioned generic 
``Love numbers''. The terms in \eqref{SpinLove} however seem to represent a new type of effects that would 
be relevant only for spinning objects, 
preceded by a new type of coefficients in \eqref{scalingWilsonS}. 
Yet, at this high perturbative order, in which quadratic-in-spin effects have already entered at many 
subleading corrections, these seemingly new terms may not correspond to a real physical effect, but rather 
could be possibly removed by virtue of subleading equations of motion, or more formally via some 
complicated subleading redefinitions of the field and worldline variables. Such spurious terms in the theory are referred to as ``redundant operators'' in EFT parlance, and simply vanish from physical observables \cite{Levi:2018nxp}. Indeed, due to the high complexity of the present sectors it is virtually impossible to identify such a redundancy by any means other than actually computing the total observables, and therefore we must press on with the full-scale evaluation of the sectors to discover the nature of the terms in \eqref{SpinLove}.

\paragraph{From EFT formulation through to observables.} 

To proceed towards the physics of the present sectors, we need to obtain first the effective action of 
the quadratic-in-spin interactions via an evaluation of the diagrammatic expansion of the two-particle 
action in \eqref{2ptact} in terms of Feynman graphs. To that end, we build on the \texttt{EFTofPNG} -- a 
unique public code for Feynman computation in PN theory \cite{Levi:2017kzq,Levi:2018stw}. First we need to 
extend the code to generate the required Feynman rules in a generic number of spatial dimensions $d$, 
including a recursive implementation of the gauge of the rotational variables, where we use the ``canonical 
gauge'' which was introduced in \cite{Levi:2015msa}. Notably, there are now spin couplings up to quadratic 
order in the curvature, in addition to a proliferation of time derivatives on the spin couplings, and from 
the gravitational self-interaction. 

The Feynman graphs that contribute to these sectors are then generated through another extension of 
\texttt{EFTofPNG}, and we find that there are 1122 graphs that make up the complete 
next-to-next-to-next-to-leading order (N$^3$LO) quadratic-in-spin sectors, 
and are of the highest complexity ever tackled in sectors with spins as yet.
This large volume of intricate graphs also carries a higher tensorial load than ever, due to the derivative 
coupling of spins, on top of the high PN order. It was thus essential to streamline new code for the 
projection of integrals, due to the high rank of their numerators \cite{Passarino:1978jh}, and for 
algorithmic integration by parts (IBP) to reduce the integrals that show up to basic master 
integrals \cite{Laporta:2001dd, Smirnov:2012gma,Levi:2020uwu}. In general, to verify the reliability of our 
new codes, we carried out independent code development in parallel, and new results were compared at 
crucial check points along the elaborate derivation.  

\begin{figure}[t]
\centering
\includegraphics[scale=0.5]{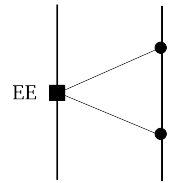}
\caption{The unique graph at the N$^3$LO quadratic-in-spin sectors, which arises from the 
quadratic-in-curvature coupling that is also quadratic in the spin, and is preceded by an unknown 
coefficient, see \eqref{SpinLove}, \eqref{scalingWilsonS}, \eqref{uniquegraph}. The coupling is denoted 
by a black square labeled EE.}
\label{n3los2e2}
\end{figure}

An expansion of the terms in \eqref{SpinLove}, \eqref{scalingWilsonS}, reveals that only the term with the 
electric curvature component actually enters at this PN order. This term gives rise to a unique graph of a 
two-graviton exchange depicted in Fig.~\ref{n3los2e2} \cite{Vermaseren:1994je,Binosi:2008ig}, whose value 
is given by
\be
\label{uniquegraph}
\text{Fig.~1}= 
-\frac{1}{2}C_{1(\text{E}^2\text{S}^2)} \frac{G^4 m_1m_2^2}{r^6}
\left[S_1^2+3\big(\vec{S}_1\cdot\vec{n}\big)^2\right].
\ee

As noted the Feynman graphs are evaluated using dimensional regularization with the $\overline{\text{MS}}$ 
scheme in \eqref{msbar}, and similar to \cite{Levi:2020uwu} the values of the individual graphs contain 
poles in the dimensional parameter, $\epsilon_d \equiv d-3$, in conjunction with logarithms in $r/R_0$. 
However, whereas in the piece 
which was approached in \cite{Levi:2020uwu}, all the poles and logarithms conspired to cancel out from 
the total sum, summing over all the graphs in our sectors does leave such divergent and logarithmic terms, 
which do not cancel out from the total sum. 
In fact, the expansion of results in terms of $\epsilon_d$ that is required here, makes for one of the most 
computationally demanding tasks in the evaluation of the present sectors. 

Summing up all of the graphs' values we get an initial action with many large pieces that contain terms with
higher-order time derivatives -- up to 6th order in the total number of derivatives. To remove these terms, 
we need to extend the formulation of a rigorous procedure that was uniquely introduced in 
\cite{Levi:2014sba}, for subleading redefinitions of both position and rotational variables. This elaborate 
procedure is critical here since such redefinitions, e.g.~$\vec{x}\to \vec{x}+\Delta \vec{x}$, need to be 
applied beyond linear order, e.g.~beyond ${\cal{O}}(\Delta \vec{x})$, in sectors that are beyond linear in 
the spins, and at high PN orders. 
The redefinitions of rotational variables, the Lorentz matrices $\Lambda^{ij}$ and the spins $S^{ij}$, 
are parametrized by $\omega^{ij}$, the anti-symmetric generator of rotations on the Lorentz matrices,
\be
\Lambda^{ij} \equiv \Lambda^{ik} \left(e^{\omega}\right)^{kj}.
\ee
We then require here for the first time redefinitions of position which depend on the spins, in 
addition to redefinitions of the rotational variables, that both scale as: 
\be
|\Delta \vec{x}| \sim \epsilon_d^{-1}, \quad |\omega_{ij}| \sim \epsilon_d^{-1}.
\ee
Further, to reduce the action to a standard action without higher-order time derivatives, all sectors up to 
quadratic-in-spin and to this PN order should be treated consistently, i.e.~from LO to N$^3$LO of 
point-mass sectors (namely Newtonian through to 3PN order), and further through spin-orbit, to 
quadratic-in-spin sectors. We get contributions to the present sectors from the application of 
redefinitions in all these sectors. Notably, further logarithmic terms arise, preceded by the 
dimensional parameter, $\epsilon_d$, 
and thus the Newtonian and the LO quadratic-in-spin interactions should all be expanded here 
in $\epsilon_d$, as these yield contributions to the present sectors.

Altogether after this arduous reduction procedure, all the divergent and logarithmic terms vanish when 
going to observables or they can be removed from the reduced action already through the addition of a total 
time derivative as we also verified. At this stage the transition to a generic Hamiltonian, as 
well as to various observables in simplified binary configurations, of e.g.~circular orbits and aligned 
spins, is straightforward \cite{Levi:2014sba,Levi:2015msa,Levi:2015uxa}. 
The total binding energy of the N$^3$LO quadratic-in-spin interactions can be written as the following sum of pieces:
\begin{align}
\label{totalSquad}
(e)_{\text{S$^2$}}^{\text{N$^3$LO}} =  
  (e)_{\text{S$_1$S$_2$}}^{\text{N$^3$LO}} 
+ &\bigg[(e)_{\text{S$_1^2$}}^{\text{N$^3$LO}} 
+ (e)_{\text{C$_{1(ES^2)}$S$_1^2$}}^{\text{N$^3$LO}}\nn\\
& + (e)_{\text{C$_{1(E^2S^2)}$S$_1^2$}}^{\text{N$^3$LO}}  
+ \left( 1\leftrightarrow2 \right) \bigg].
\end{align}
As a function of the orbital frequency parameter $x$, these various pieces are given by
\begin{widetext}
\begin{align}
\label{es1s2}
(e&)^{\text{N}^3\text{LO}}_{\text{S}_1 \text{S}_2} (x) = \tilde{S}_1 \tilde{S}_2 x^6 \nu 
\left[ \frac{243}{16} -\left( \frac{2107}{16}  - \frac{123}{32} \pi ^2 \right) \nu 
+ \frac{147}{8} \nu^2 + \frac{13}{16} \nu^3 \right],\\\nn\\
\label{es2}
(e&)^{\text{N}^3\text{LO}}_{\text{S}_1^2} (x)  = \tilde{S}_1^2 x^6 \nu 
\left[ \left(\frac{1947}{112} \nu -\frac{48357}{560} \nu^2 + \frac{159}{16} \nu ^3 \right)  
-q^{-1} \left(\frac{243}{16} - \left( \frac{747}{16} - \frac{189 }{2048} \pi ^2 \right) \nu
 + \frac{13731}{280} \nu ^2- \frac{153}{16} \nu ^3 \right)\right],\\
\label{ecs2}
(e&)^{\text{N}^3\text{LO}}_{\text{C$_{1(ES^2)}$S$_1^2$}} (x)  = C_{1(\text{ES}^2)} \tilde{S}_1^2 
x^6 \nu \left[ \left(\frac{789}{28} \nu - \frac{156}{7} \nu^2 + \frac{5}{8} \nu ^3\right) 
+ q^{-1} 
\left( \frac{405}{32} - \left(\frac{2389}{32}-\frac{3747}{2048} \pi^2\right) \nu 
-\frac{555}{56} \nu^2 + \frac{21}{32} \nu^3\right) \right],
\end{align}
\end{widetext}
\bea\label{es2E2}
(e)^{\text{N}^3\text{LO}}_{\text{C$_{1(E^2S^2)}$S$_1^2$}} (x) = 
-\frac{3}{2} C_{1(\text{E}^2\text{S}^2)}\tilde{S}_1^2 x^6 \nu 
\left[ \nu^2 (1+q^{-1}) \right],\qquad
\eea
where all definitions and notations are identical to \cite{Levi:2014sba}. 
These results were obtained by going from our most general new results to the simplified specific 
configuration of circular orbits with aligned spins. 
The simplest piece is linear in the individual spins as shown in \eqref{es1s2}, and is in agreement 
with \cite{Antonelli:2020ybz}, who obtained it only within a limited treatment of the simplified specific 
case of circular orbits with aligned spins in the center-of-mass frame, unlike our most generic treatment. 

Importantly we also find that there is a new type of contribution in \eqref{es2E2} due to the new term from 
\eqref{SpinLove}, \eqref{scalingWilsonS}. This means that this new term is not a ``redundant operator'' in 
the EFT, but rather represents a real new physical effect which is \textit{unique to spinning objects}. It 
can be verified that \eqref{es2E2} is always negative, and thus it increases the binding energy of the 
compact binary, similar to the effects linked with the long-known ``Love numbers'' in \eqref{ordinaryLove}. 
Thus, we also discovered here a new ``Spin Love number'', which is \textit{unique to spinning objects}. 

We also find the following pieces for the gauge-invariant relation of the binding energy to the angular 
momentum:
\begin{widetext}
\be
(e)^{\text{N}^3\text{LO}}_{\text{S}_1 \text{S}_2} (\tilde{L}) = 
- \tilde{S}_1 \tilde{S}_2 \frac{\nu}{\tilde{L}^{12}} 
\left[ \frac{102897}{16} - \left( \frac{31653}{32} - \frac{369}{32} \pi^2 \right) \nu 
- \frac{579}{16} \nu^2 + \frac{209}{64} \nu^3  \right], 
\ee
\begin{align}
(e)^{\text{N}^3\text{LO}}_{ \text{S}_1^2} (\tilde{L}) = 
- \tilde{S}_1^2\frac{\nu}{\tilde{L}^{12}} & \left[
\left(\frac{2117357}{896}  \nu  - \frac{714891}{2240} \nu^2 + \frac{201}{128} \nu^3 \right) \right.\nn\\
& \left.
+\,q^{-1} \left(\frac{211653}{128} + \left( \frac{21195}{16} - \frac{63}{2048} \pi^2  \right) \nu
- \frac{167739}{560} \nu^2 + \frac{3}{32} \nu^3 \right) \right],\\
(e)^{\text{N}^3\text{LO}}_{\text{C$_{1(ES^2)}$S$_1^2$}} (\tilde{L}) = 
- C_{1(\text{ES}^2)} \tilde{S}_1^2\frac{\nu}{\tilde{L}^{12}} & \left[ 
\left(\frac{2593}{14} \nu  - \frac{319}{28} \nu^2 + \frac{3}{8} \nu^3 \right) \right.\nn\\
& \left.
+\,q^{-1} \left(\frac{16065}{32} - \left( \frac{3061}{32} - \frac{11745}{2048} \pi^2  \right) \nu
- \frac{313}{28} \nu^2 + \frac{17}{32} \nu^3 \right) \right],
\end{align}
\end{widetext}
\bea
(e)^{\text{N}^3\text{LO}}_{\text{C$_{1(E^2S^2)}$S$_1^2$}} (\tilde{L}) = 
\frac{1}{2} C_{1(\text{E}^2\text{S}^2)}\tilde{S}_1^2 \frac{\nu}{\tilde{L}^{12}}  
\left[  \nu^2 \left( 1 + q^{-1} \right) \right].\qquad
\eea
These relations provide a useful tool for evaluating different analytic and numerical 
descriptions of the binary dynamics, see \cite{Levi:2014sba} and references therein.

\paragraph{State of the art and new physics.} 

The EFT of spinning gravitating objects introduced in \cite{Levi:2015msa} provides a self-contained 
framework that allows for significant formal and technical extensions which in turn enable to push the 
precision frontier, as demonstrated in this letter. Our framework handles generic compact binaries, and 
provides a host of useful mathematical and observable quantities, which are not limited to simplified 
specific binary configurations, such as circular orbits or no eccentricity, and the aligned-spins cases. 
In this letter we pushed the state of the art of the conservative dynamics at the 5PN order in these most 
generic settings for all quadratic-in-spin sectors. The push in PN accuracy already greatly improves our 
ability to learn on the fundamental physics that is encrypted in GW data. Yet, the state of the art 
accomplished in this letter is even more crucial as it handles the real-world spinning case, and moreover 
goes beyond linear order in spins, namely to finite-size effects that provide unique information on 
gravity and QCD.    

Our framework here also enabled to discover a new type of physical effect that is \textit{unique to 
spinning objects}, with a new ``Spin Love number''. The new effect enters at the 5PN interaction, and 
binding energy in \eqref{totalSquad}, similar to the spinless terms in \eqref{ordinaryLove} with 
coefficients in \eqref{scalingWilson} that correspond to the long-known ``Love numbers''. As the 5PN 
frontier has been recently approached, there has been a surge of studies on these long-known Love numbers 
-- for rotating BHs and NSs 
\cite{LeTiec:2020spy,LeTiec:2020bos,Chia:2020yla,Charalambous:2021mea,Charalambous:2021kcz,
Castro:2021wyc}. These studies by various groups arrived at several new intriguing findings and insights did not reach unanimity in whether or not these Love numbers vanish for rotating black holes, and seem to be far from being concluded.
It is thus vital to thoroughly tackle this challenging and rich line of study on these coefficients for 
generic compact objects in various approaches, and for various general theories (including in generic 
dimensions, see e.g.~\cite{Kol:2011vg}), as this is bound to uncover new physics. It remains for future 
analytical and numerical studies, and analysis of GW data, to also uncover the unique new physics, that is 
encapsulated in the new coefficients or ``Spin Love numbers'' discovered here in this letter. 


\begin{acknowledgments}
\paragraph{Acknowledgments.}
We thank Johannes Bl\"{u}mlein and Alex Edison for pleasant discussions.
J-WK was supported by the Science and Technology Facilities Council (STFC) 
consolidated grant ST/T000686/1 \textit{“Amplitudes, Strings and Duality”}.
ML received funding from the European Union's Horizon 2020 research and 
innovation programme under the Marie Sk{\l}odowska-Curie grant 847523, 
and is supported by the STFC Rutherford grant ST/V003895/1 
\textit{``Harnessing QFT for Gravity''}. 
ZY is supported by the Knut and Alice Wallenberg Foundation under grants 
KAW 2018.0116 and KAW 2018.0162.
\end{acknowledgments}

\bibliography{gwbibtex}

\end{document}